\documentclass[english]{article}
\usepackage{lmodern}

\usepackage[T1]{fontenc}
\usepackage[latin9]{inputenc}
\usepackage{float}
\usepackage{booktabs}
\usepackage{mathrsfs}
\usepackage{amsmath}
\usepackage{amssymb}
\usepackage{graphicx}

\makeatletter

\providecommand{\tabularnewline}{\\}

\newcommand{\lyxaddress}[1]{
\par {\raggedright #1
\vspace{1.4em}
\noindent\par}
}

\usepackage{mathrsfs}   
\usepackage{slashed}     
\usepackage{bbold}  
\usepackage{url}
\usepackage{graphicx}
\usepackage[colorlinks=true,linkcolor=redLinks,citecolor=greenLinks,urlcolor=redLinks, pdfborder={0 0 1}]{hyperref}
\usepackage{xcolor}
\usepackage{framed}
\usepackage[numbers,sort&compress]{natbib}
\usepackage{amsmath}

\allowdisplaybreaks

\colorlet{shadecolor}{gray!15}

\definecolor{greenLinks}{rgb}{0, 0.6, 0} 
\definecolor{blueLinks}{rgb}{0, 0, 0.6}
\definecolor{redLinks}{rgb}{0.6, 0, 0}
\definecolor{tempText}{rgb}{0.55, 0.10,0.67}
\definecolor{eprintLinks}{rgb}{0.4, 0.4, 0.4}
\definecolor{journalLinks}{rgb}{0.6, 0, 0}

\newcommand{\MYhref}[3][redLinks]{\href{#2}{\color{#1}{#3}}}%

\usepackage{multirow}
\let\orig@Hy@EveryPageAnchor\Hy@EveryPageAnchor
\def\Hy@EveryPageAnchor{%
    \begingroup
    \hypersetup{pdfview=Fit}%
    \orig@Hy@EveryPageAnchor
    \endgroup
}

\let\oldFootnote\footnote
\newcommand\nextToken\relax

\renewcommand\footnote[1]{%
    \oldFootnote{#1}\futurelet\nextToken\isFootnote}

\newcommand\isFootnote{%
    \ifx\footnote\nextToken\textsuperscript{,}\fi}

\definecolor{myPurple}{RGB}{128,0,182}

\let\oldoverline\overline
\renewcommand{\overline}[1]{\mkern 1.5mu\oldoverline{\mkern-1.5mu#1\mkern-1.5mu}\mkern 1.5mu}

\makeatother

\usepackage{babel}
\begin{document}

\title{{\Large{}\vspace{-1.0cm}} \hfill {\normalsize{}IFIC/18-16} \\*[10mm]
{\huge{}$\Delta L \ge 4$ processes}{\Large{}
\vspace{0.5cm}}}

\author{{\Large{}Renato M. Fonseca}\thanks{E-mail: renato.fonseca@ific.uv.es; Website: \href{http://renatofonseca.net/}{renatofonseca.net}	 }
  {\Large{}, Martin Hirsch}\thanks{E-mail: mahirsch@ific.uv.es} \date{}}

\maketitle

\lyxaddress{\begin{center}
{\Large{}\vspace{-0.5cm}}AHEP Group, Instituto de Física Corpuscular,
C.S.I.C./Universitat de València\\
Edifício de Institutos de Paterna, Apartado 22085, E--46071 València,
Spain
\par\end{center}}

\begin{center}
\today
\par\end{center}
\begin{abstract}

We discuss the experimental prospects for observing processes which
violate lepton number ($\Delta L$) in four units (or more).  First, we reconsider
neutrinoless quadruple beta decay, deriving a model independent and
very conservative lower limit on its half-life of the order of
$10^{41}$ ys for $^{150}$Nd. This renders quadruple beta decay
unobservable for any feasible experiment. We then turn
to a more general discussion of different possible low-energy
processes with values of $\Delta L \ge 4$. A simple operator analysis
leads to rather pessimistic conclusions about the observability at
low-energy experiments in all cases we study. However, the situation
looks much brighter for accelerator experiments. For two
example models with $\Delta L=4$ and another one with $\Delta L=5$, we show how the LHC
or a hypothetical future pp-collider, such as the FCC, could probe
multi-lepton number violating operators at the TeV scale.  \\ \\ \\ \\ \\

\end{abstract}
\newpage{}

\section{Introduction}

So far, no lepton ($L$) nor baryon ($B$) number violating process has been
observed experimentally. However, there are good reasons to believe
that neither of these quantities are actually conserved. In fact, even within the
standard model (SM), non-perturbative effects such as the sphaleron
\cite{Klinkhamer:1984di} violate both $B$ and $L$. More
phenomenologically, also the observed baryon asymmetry of the
universe points to the existence of $B$ violation at some point in its
early history.

From the viewpoint of standard model effective theory, one can build
non-renormalizable operators which violate $B$ and $L$
\cite{Weinberg:1979sa,Weinberg:1980bf}. The lowest dimensional
operator of this kind, the Weinberg operator, appears at dimension 5 ($d=5$) and
violates $L$ by two units. This operator generates Majorana neutrino mass terms
which can be experimentally probed by the observation of neutrinoless double beta decay,
$0\nu\beta\beta$: $\left(A,Z\right)\rightarrow\left(A,Z \pm
2\right)+2e^{\mp}$ (for recent reviews see for example
\cite{Avignone:2007fu,Deppisch:2012nb}).  Next, at $d=6$, one finds
$\Delta B=\Delta L=1$ operators. These operators cause proton decay in
modes such as the famous $p \to e^+ \pi^0$.  This and other two-body
nucleon decays are well-known to arise in a variety of models, most
notably in Grand Unified Theories --- see
\cite{Langacker:1980js,Nath:2006ut} and references contained therein.

The gauge structure of the SM and its field content is such that $\Delta L= 2n + \Delta B$ for
all non-renormalizable operators ($n$ being an integer). Thus, for example, even $\Delta L$ is associated to
even $\Delta B$, so no proton decay mode with $\Delta L=2$ can
exist. However, starting at $d=9$ one finds $\Delta L=3$ operators, associated to $\Delta B=1$. Also, operators relevant for $\Delta B=2$ processes, such as
neutron-antineutron oscillations, appear first at $d=9$. One would naively assume that the rates
for $\Delta L \ge 3$ (or also $\Delta B \ge 2$) processes are
necessarily much smaller than those corresponding to the lower dimensional non-renormalizable operators. However, this may not be the case, and in fact it is possible that $\Delta B,\Delta L=1$ processes are forbidden altogether.  This is
exactly what happens in the Standard Model, since sphalerons are $\Delta
B=\Delta L=3$ transitions (thus, sphalerons cannot destroy
protons).\footnote{It is interesting to note that it has long been
	believed that sphaleron transitions are unobservable
	experimentally. However, some recent papers on sphaleron rates at
	accelerators have come to much more optimistic conclusions
	\cite{Tye:2015tva,Ellis:2016ast,Tye:2017hfv}.}

There is also the possibility that beyond the SM there exists some (so
far unknown) symmetry such that lepton number and/or baryon number can
only be created or destroyed in larger multiples. For $\Delta L=3$
this has been recently discussed in \cite{Fonseca:2018ehk}, see also
\cite{Hambye:2017qix}.  In that case, for example, standard proton
decay modes are absent and one is left with $p\rightarrow
e^{+}\overline{\nu}\overline{\nu}$, $\pi^{0}e^{+}\overline{\nu}\overline{\nu}$,
$e^{-}\nu\nu\pi^{+}\pi^{+}$ and, more interestingly,
$e^{+}e^{+}e^{+}\pi^{-}\pi^{-}$. As noted above, these processes are
induced by $d=9$ or higher operators, which implies that the proton
decay rate is suppressed by many powers of the new physics scale
$\Lambda$.  Consequently, one can have $\Lambda\sim\textrm{TeV}$,
making it possible for colliders to probe this
hypothesis \cite{Fonseca:2018ehk}.

In this paper, we will discuss $\Delta L=4$ processes (addressing also
the possibility of having $\Delta L \ge 5$), and analyse whether any of
these processes can possibly be observed in the foreseeable
future. Specific example models will be
constructed where, due to the presence of some symmetry,
operators involving less leptons are forbidden.

We start by noting that all $\Delta L=4$ operators must have dimension
greater or equal 10 (see table \ref{tab:ExaOp} for examples).  Note that at $d=10$ there is just one unique $\Delta
L=4$ operator.\footnote{We use the word \textit{operator} to denote a combination of fields, independently of the number of contractions. In the case of $L_iL_jL_kL_lHHHH$ (subscripts denote the lepton flavors), there are two contractions: the square of the Weinberg operator, $\mathcal{O}_{ijkl}=\mathcal{O}_{ij}^{W}\mathcal{O}_{kl}^{W}$, and another one $\mathcal{O}^\prime_{ijkl}$. Nevertheless, it is not necessary to include in the Lagrangian this last one because it is related to the $\mathcal{O}$ contraction with the $ijkl$ indices permuted. Furthermore, note that there are several identities among the entries of the tensor $\mathcal{O}_{ijkl}$, so a total of six couplings/numbers encode all possible $L_iL_jL_kL_lHHHH$ interactions (for example, if three or more lepton indices are the same, the operator is identically zero).}  At $d=12$ there are already 8 different operators
(including derivatives), so we show only some examples for $d\ge
12$.\footnote{The complete list up to $d=22$, compiled with the Sym2Int program \cite{Fonseca:2017lem}, can be found at \href{http://renatofonseca.net/dl4OpList.php}{renatofonseca.net/dl4OpList.php}.}
Some operators require more than one generation of fermions, such as
the operator at $d=10$. Adding two derivatives to this operator, a
very similar operator can be realized with just one generation of
leptons at $d=12$.  We will discuss these two operators in more detail
in sections \ref{sec:0Nu4Beta} and \ref{sec:Colliders}.  The second
example at $d=12$ simply exchanges one Higgs from the operator at
$d=10$ for two quarks ${\bar Q}{\bar u^c}$, equivalent to a Yukawa
interaction. In this way, higher dimensional operators with less Higgs
fields can be constructed. This is important, as will be seen in
sections \ref{sec:0Nu4Beta} and \ref{sec:OtherLowEnergyProcesses}, for
experiments at low energies.  Also at $d=12$ we can find an operator
involving only fermions (third example in
table \ref{tab:ExaOp}). However, for this particular operator the final
state in any process will always involve neutrinos, making it
impossible to tag the lepton number experimentally.

We also show in the table two examples of $\Delta L=4$ operators with $d=15$, both of which violate baryon number.
The first one has two
$L$ fields, thus the $SU(2)_L$ contractions again lead to final states
involving a neutrino. We will be more interested in the other operator, with four charged leptons.
In fact, its realization/decomposition necessarily involves new coloured fields which can be
searched at the LHC, as we will point out in section
\ref{sec:Colliders}.

\begin{center}
	\begin{table}[tbph]
		\begin{centering}
			\begin{tabular}{ccc}
				\toprule 
				$\Delta B$ & Operator dimension & Example operators 
				\tabularnewline
				\midrule
				0 & 10 & $L L L L H H H H$
				\tabularnewline
				0 & 12 & $\partial\partial L L L L H H H H$,
				${\bar Q}{\bar u^c} LLLL HHH$, 
				${\bar u^c}{\bar u^c}d^c d^c L L L L$
				\tabularnewline
				0 & 14 & $\partial\partial{\bar u^c}{\bar u^c}d^cd^c LLLL$,  
				${\bar u^c}d^c {\bar e^c}{\bar e^c}{\bar e^c}LHHHHH$
				\tabularnewline
				2 & 15 & 
				${\bar u^c}{\bar u^c}{\bar u^c}{\bar u^c}{\bar u^c}{\bar d^c}
				{\bar e^c}{\bar e^c}LL$, ${\bar u^c}{\bar u^c}{\bar u^c}{\bar u^c}{\bar u^c}{\bar u^c}
				{\bar e^c}{\bar e^c}{\bar e^c}{\bar e^c}$
				\tabularnewline
				\bottomrule
			\end{tabular}
			\par\end{centering}
		
		\protect\caption{\label{tab:ExaOp}Some examples of $\Delta L=4$
			operators. Note that the lowest dimensional operator at $d=10$ is
			unique. All odd-dimensional operators with $d<23$ and $\Delta L=4$ violate also $B$, and the smallest ones appear at $d=15$. }
	\end{table}
	
	\par\end{center}

As far as we know, the only $\Delta L=4$ process treated in some detail in
the literature is neutrinoless quadruple beta
decay ($0\nu4\beta$), having been discussed for the first time in
\cite{Heeck:2013rpa}.  A simple power counting suggests that the decay
rate associated to $0\nu4\beta$ will be extremely small if all
particles which mediate this process are heavy \cite{Hirsch:2017col}.
However, the analysis in \cite{Heeck:2013rpa,Hirsch:2017col} leaves
open the possibility of having an observable $0\nu4\beta$ decay rate,
if several of the mediator particles are very light (here ``light''
implies masses of the order of the nuclear Fermi scale, i.e. ${\cal O}(0.1)$
GeV). Thus, in section \ref{sec:0Nu4Beta} we
calculate a very conservative lower limit on the $0\nu4\beta$ decay lifetime, 
based on collider searches for charged bosons. We find that this minimum lifetime is around 20 orders
of magnitude larger than the current experimental limit
\cite{Arnold:2017bnh}, rendering $0\nu4\beta$ virtually unobservable.

In section \ref{sec:OtherLowEnergyProcesses} we extend the
discussion to other $\Delta L=4,5,6,\cdots$ low-energy processes. In all cases our simple rate estimate is far below
experimental sensibilities. The most optimistic case, dinucleon decay
of the form $(A,Z) \to (A-2,Z-2) + 2 \pi^- + 4 e^+$, is expected to be at least some $8$
orders of magnitude beyond the current Super-Kamiokande sensitivity. 
Thus it seems impossible for low-energy experiments to test lepton number
violation in 4 or more units.

However, the prospects of observing $\Delta L\ge 4$ processes at
colliders are good, as we discuss in section \ref{sec:Colliders}.
Indeed, because operator dimensionality becomes irrelevant at energies
comparable to the new physics scale, the LHC is in a good position to
probe the creation or destruction of groups of 4 or more charged
leptons. We construct two example models for $\Delta L=4$ (and
one for $\Delta L=5$) and calculate production cross sections for
the different particles in these models at pp-colliders (the
LHC and a hypothetical $\sqrt{s}=100$ TeV collider). From these
cross sections we derive lepton number violating event rates and
estimate the scales up to which pp-colliders can test such kind
of models. We then end the paper with a summary.

\section{\label{sec:0Nu4Beta}Minimum lifetime of neutrinoless quadruple beta
	decay}

The possibility of observing $\Delta L=4$ processes via neutrinoless
quadruple beta decay ($0\nu 4\beta$) was put forward in
\cite{Heeck:2013rpa}. Three candidate nuclei for the decay
$\left(A,Z\right)\rightarrow\left(A,Z+4\right)+4e^{-}$ were
identified: by far the most promising one is $^{150}\textrm{Nd}$,
which could decay into $^{150}\textrm{Gd}$ plus four electrons with a
total kinetic energy of $Q=2.08$ MeV.\footnote{The isotope
	$_{50}^{126}\textrm{Sn}$ may decay into
	$_{54}^{126}\textrm{Xe}+4e^{-}$, with $Q=3.13$ MeV
	\cite{Audi:2017asy}. However, it can also undergo single beta decay,
	with a half-life of $2.3\times10^{5}$ years.} With 36.6g of the
isotope, the NEMO-3 detector is well suited to measure this decay
provided that it happens at a reasonable rate.  Recently, based on an
exposure of 0.19 kg$\cdot$y of $^{150}\textrm{Nd}$, the collaboration
reported a lower limit of $\left(1.1-3.2\right)\times10^{21}\textrm{
	y}$ on the half-life for this particular process, at a 90\%
confidence level \cite{Arnold:2017bnh}. The range in this number is
explained mostly by the fact that, so far, no reliable theoretical
calculation of the single electron spectra has been done for this
process.  Nevertheless, as can be seen, the result is fairly
insensitive to such details.

\begin{figure}[H]
	\begin{centering}
		\includegraphics[scale=1.2]{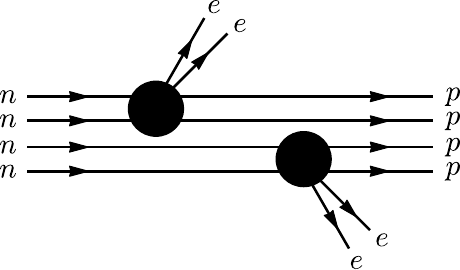}
		\par\end{centering}
	
	\protect\caption{\label{fig:trivial-quadruple-beta-decay}Quadruple
		beta decay induced by two virtual double beta decays. This
		contribution to the decay rate is necessarily very small, given the
		current limits on the $0\nu2\beta$ decay life-time of various
		nuclei. }
\end{figure}

Given the non-observation of $0\nu2\beta$ decay events so far, one
has to wonder if $0\nu4\beta$ decays will ever be observed. If the
only contribution to the latter process is the conversion of a pair
of neutrons into protons and two electrons, twice repeated (see fig.
\ref{fig:trivial-quadruple-beta-decay}), then one would expect the
approximate relation
\begin{align}
\tau_{0\nu4\beta} & \sim\left(\frac{\tau_{0\nu2\beta}}{10^{26}\textrm{y}}\right)^{2}\left(\frac{q}{100\textrm{ MeV}}\right)^{2}\left(\frac{\textrm{MeV}}{Q}\right)10^{88}\textrm{y}
\end{align}
between the lifetimes of double and quadruple beta decay without
neutrinos.\footnote{This naive estimate is based on a simple
	dimensional analysis, assuming that both processes (involving
	potential different parent nuclei) have the same kinetic energy
	$Q$. This coarse assumption inevitably introduces a large error in
	the result which, nevertheless, is of no material significance in
	the face of such large lifetimes.} In this expression $q\sim100$ MeV
stands for the typical momentum transfer in a nucleus. Using the
experimental lower limit on $\tau_{0\nu2\beta}$ for various nuclei of
the order of $10^{26}$ years
\cite{KamLAND-Zen:2016pfg,Agostini:2018tnm}, we can extract the lower
bound $\tau_{0\nu4\beta}\gtrsim10^{88}$ years.

However, this estimate assumes that the main contribution to
neutrinoless quadruple beta decay comes from two virtual double beta
decays. This does not need to be the case; indeed, as previously
mentioned, it is possible to forbid entirely $\Delta L=2$ processes
and still have those with $\Delta L=4$.

In the following we will argue that an important constraint on
neutrinoless quadruple beta decay can be derived using data from
collider searches of charged bosons. However, before proceeding, let
us consider first what is the expected value of the $0\nu4\beta$ decay
rate from simple power counting (see also
\cite{Heeck:2013rpa,Hirsch:2017col}).  We start by noting that there
are 12 fermions involved, so after electroweak symmetry breaking the
relevant operator has dimension 18:
\begin{align}
\mathscr{O}_{0\nu4\beta} & =\frac{\kappa}{\prod_{i=1}^{14}\Lambda_{i}}\overline{u}\overline{u}\overline{u}\overline{u}dddd\overline{e}\overline{e}\overline{e}\overline{e}\,.\label{eq:operator_after_EWSB}
\end{align}
Here, $\kappa$ is just some unspecified dimensionless coefficient.  On
the other hand, in the final state there are four electrons and a
nucleus which stays essentially at rest, so the $0\nu4\beta$ decay
rate will depend on the $(3\times4-1)$th power of the available
kinetic energy $Q$. With these considerations, inserting some
numerical factors for the multi-body kinematics, we obtain the formula
\begin{align}
\tau_{0\nu4\beta} & \sim\kappa^{-2}\left(\frac{Q}{\textrm{MeV}}\right)^{-11}\left(\frac{q}{100\textrm{ MeV}}\right)^{-18}\prod_{i=1}^{14}\left(\frac{\Lambda_{i}}{\textrm{TeV}}\right)^{2}10^{110}\textrm{y}\, .\label{eq:3}
\end{align}
Clearly, this is a very rough estimate for the life-time of the
process, with an uncertainty of a few orders of magnitude.
Nevertheless, given the largeness of the numbers involved, it is good
enough for the following discussion.

To obtain the lowest possible $\tau_{0\nu4\beta}$, ideally one would
need large, i.e. order ${\cal O}(1)$, couplings ($\kappa\sim1$) and
light mediator masses ($\Lambda_{i}\ll\textrm{TeV}$). Note that for
very light particles, their mass becomes irrelevant when compared to
the typical momentum transfer in the nucleus. Hence the best case
scenario is $\Lambda_{i}^{min}\sim q$.  One can see easily that, in
the limit where all $\Lambda_{i}$ are of the order of 1 GeV or lower,
it becomes conceivable to have
$\tau_{0\nu4\beta}\left(^{150}\textrm{Nd}\right)$ smaller than
$10^{26}$ years.

However, we will now show that the $0\nu4\beta$ decay diagram contains
at least 4 propagators of charged bosons (scalar or vector). The
lightest particle, that can play this role, is the $W$ boson, hence
the most optimistic scenario achievable is
$\prod_{i=1}^{14}\Lambda_{i}=m_{W}^{8}q^{6}$, which translates into
\begin{align}
\tau_{0\nu4\beta}^{\textrm{Min}}\left(^{150}\textrm{Nd}\right) & \sim10^{41}\textrm{y}\,.\label{eq:4}
\end{align}
This corresponds to roughly one $0\nu4\beta$ decay per year for a
mass of $\sim10^{17}$ kg of neodymium, hence the observation of this
process would be extremely challenging even in the most optimistic
scenario.

Independent on any concrete model, four or more of charged bosons
propagators are needed for the following reason. Conservation of
fermion number implies that the 12 external fermions which make up the
$0\nu4\beta$ operator must be arranged in six currents $J_{i}$. Out of
the six possible pairings ($\overline{u}\overline{u}$,
$\overline{u}d$, $\overline{u}\overline{e}$, $dd$, $d\overline{e}$ and
$\overline{e}\overline{e}$) none is electrically neutral. 
Thus all six $J_{i}$ currents must exchange charge through scalar or
vector bosons. (It is also conceivable that this charge exchange
occurs through some intermediary vertex, yet this scenario does not
lead to a minimal number of internal propagators.)

It is straightforward to see that the most economical setup is the
one, where pairs of currents $J_{i}$ with opposite electric charged
are connected by a single charged boson, in which case only 3 such
propagators are needed.
However, the only fermion billinears with opposite charges are $dd$
and $d\overline{e}$, and clearly one does not obtain the $0\nu4\beta$
operator with three copies of these fermions. Hence, a minimum of
4 charged bosons are needed and the lower limit in eq. (\ref{eq:3})
applies.

The six possible currents need to be coupled to scalars or vectors.
These are charged or doubly charged particles, leptoquarks or 
diquarks. If one were to construct a loop model for $0\nu 4\beta$
decay and introduce also some new exotic fermions, more exotic
scalars/vectors could, in principle, also appear. All of these states,
however, necessarily couple to standard model fermions and thus can be
searched at accelerators such as LEP and LHC. Again, considering the
large numbers involved in eqs. (\ref{eq:3}) and (\ref{eq:4}), a very
rough argument suffices for our purpose.  Thus, we only quote that LEP
data rules out any electrically charged boson, decaying to SM
fermions, below roughly 100 GeV \cite{Patrignani:2016xqp},

The lower bound in eq. (\ref{eq:4}) on the neutrinoless quadruple
beta decay lifetime is therefore unavoidable. However, it is also very
conservative. If one were to construct lower bounds individually for
the 6 possible currents, (much) larger limits could be derived for the
different individual cases. Instead, we will now try to see, whether
it is actually possible to approach this bound, by considering some
particularly promising models.

\begin{figure}[H]
	\begin{centering}
		\includegraphics[scale=1.2]{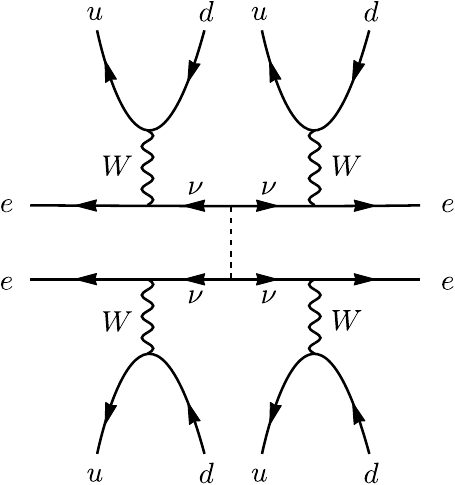}
		\par\end{centering}
	
	\protect\caption{\label{fig:best_diagram}Diagram for $0\nu4\beta$ decay with only
		Standard Model fermions and a new neutral scalar (first considered
		in \cite{Heeck:2013rpa}). Note that all fermions drawn are left-handed.}
\end{figure}

\begin{figure}[H]
	\begin{centering}
		\includegraphics[scale=1.2]{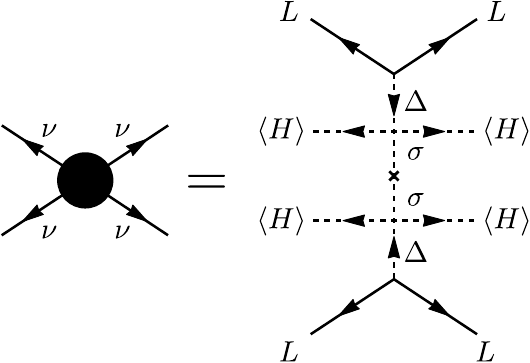}
		\par\end{centering}
	
	\protect\caption{\label{fig:LLLLHHHH}One possible way of generating a sizable four-neutrino
		interaction which is not suppressed by $m_{\nu}/\left\langle H\right\rangle $,
		introducing a scalar triplet $\Delta$ and singlet $\sigma$. If more
		than 2 neutrinos have the same flavour, the amplitude of the process
		is suppressed by a factor $\left(\textrm{momentum}/\textrm{mediator mass}\right)^{2}$.}
\end{figure}

We have established already that $0\nu4\beta$ decay is necessarily
suppressed by the heavy mass of at least 4 charged bosons and, in
order not to further reduce the decay rate, one should avoid colored
particles in internal lines.
\footnote{Bounds on coloured particles from LHC approach or exceed TeV
	masses, leading to further suppression of the rate.} It is easy to
check that, out of the 135 distinct ways of partitioning the
$0\nu4\beta$ operator in fermion billinears/currents $J_{i}$, only
$\left(\overline{u}d\right)\left(\overline{u}d\right)\left(\overline{u}d\right)\left(\overline{u}d\right)\left(\overline{e}\overline{e}\right)\left(\overline{e}\overline{e}\right)$
makes it possible to have all internal bosons colorless. It is equally
simple to arrive at the conclusion that, for this particular partition
of the $0\nu4\beta$ operator, the number of neutral and colorless internal
bosons will be minimal (just 4) if and only if the corresponding
diagram can be split into two halves, each with fermion billinears
$\left(\overline{u}d\right)\left(\overline{u}d\right)\left(\overline{e}\overline{e}\right)$,
connected by a neutral boson. A particularly interesting way of
building such diagram is the one shown in fig.
\eqref{fig:best_diagram}, where only SM fields plus a neutral scalar are
used \cite{Heeck:2013rpa}.  In the central part of the diagram, it is
clear that 4 neutrinos are created from nothing, hence the
neutrino-scalar interaction is critical for the violation of lepton
number. Under the full Standard Model symmetry, and setting aside a
complication which we shall mention later, this $\Delta L=4$ operator
must be of the form $LLLLHHHH$ (or perhaps higher
dimensional). Crucially, this effective interaction does not need to
be suppressed by the smallness of neutrino mass; indeed, the operator
$LLHH$ might even by absent. One way of generating the $LLLLHHHH$
interaction is by adding to the Standard Model an extra $SU(2)_{L}$
scalar triplet $\Delta$ with one unit of hypercharge, as well as a
real scalar singlet $\sigma$ with no gauge interactions.
The most general Lagrangian with these two
new fields violates lepton number in two units, given the simultaneous
existence of the couplings $LL\Delta$ and $HH\Delta^{*}$.
\footnote{This setup reminds of the seesaw type-II
	\cite{Schechter:1981cv}, known since a long time ago.}  It is
possible to construct a model which violates lepton number in units of
4 only, and to do so we retain just the following interactions:
\begin{align}
\mathscr{L} & =\mathscr{L}_{SM}+\left(y_{\Delta}LL\Delta+\kappa HH\Delta^{*}\sigma+\textrm{h.c.}\right)-\mu_{\sigma}^{2}\sigma^{2}+\left(\textrm{other self-conjugate terms}\right)\,.
\end{align}
This Lagrangian is invariant under the $Z_{4}(L)$ lepton-number
symmetry $\psi\rightarrow i^{L}\psi$, with
$L\left(\Delta\right)=L\left(\sigma\right)=-2$.  It is straightforward
to arrange scalar potential parameters such that only the Higgs
doublet acquires a non-zero vacuum expectation value, in which case
there is no spontaneous breaking of the $Z_{4}$ symmetry. As such, the
Weinberg operator is not generated, but $LLLLHHHH$ is (see fig.
\eqref{fig:LLLLHHHH}). Assuming without loss of generality that the
coupling $\kappa$ is real, $\sigma$ mixes only with
$\textrm{Re}\left(\Delta^{0}\right)$, resulting in two mass
eigenstates with masses $m_{1,2}$ and a mixing angle $\theta$ which
can be controlled at will. In particular, it is possible to (a) make
one of these states very light, (b) have maximal mixing $\sin\theta$,
and at the same time (c) keep all other new scalars
($\textrm{Im}\left(\Delta^{0}\right)$, $\Delta^{+}$ and $\Delta^{++}$)
arbitrarily heavy. Neutrino masses can be accounted for by introducing
right-handed neutrinos, or by the breaking of this $Z_{4}$ symmetry,
for example by introducing a very small $HH\Delta^{*}$
coupling. Likewise, the $0\nu2\beta$ decay rate can be zero (absence
of $HH\Delta^{*}$) or very small (small $\Delta L=2$ coupling).

The complication mentioned earlier is that there is no $\frac{1}{\Lambda^{2}}\nu_{e,L}\nu_{e,L}\nu_{e,L}\nu_{e,L}$
local operator: given that the spinor $\nu_{e,L}$ has two components
only, the Pauli exclusion principle forbids point interactions with
more than two $\nu_{e,L}$. Instead, the four-neutrino interaction
must be of the form
\begin{align}
\frac{1}{\Lambda^{2}}\nu_{e,L}\left(x\right)\nu_{e,L}\left(x\right)\nu_{e,L}\left(y\right)\nu_{e,L}\left(y\right) & \approx\frac{\left(y-x\right)^{2}}{\Lambda^{2}}\nu_{e,L}\left(x\right)\nu_{e,L}\left(x\right)\left[\partial\nu_{e,L}\left(x\right)\right]^{2}\,,\nonumber \\
& \approx\frac{1}{\Lambda^{4}}\nu_{e,L}\left(x\right)\nu_{e,L}\left(x\right)\left[\partial\nu_{e,L}\left(x\right)\right]^{2}\,,
\end{align}
where $\Lambda$ is a mass associated to the particle(s) mediating
the effective interaction. We used the fact that the distance between
the two interaction points $x$ and $y$ is inversely proportional
to $\Lambda$. In practice, this means that there is an extra suppression
factor $\left(\textrm{momentum}/\Lambda\right)^{2}$ due to Pauli
blocking on top of the expected $1/\Lambda^{2}$ suppression for this
4-fermion interaction\footnote{This kind of amplitude suppression can equally appear in operators
	with no fermions. For example, consider a global $SU(2)$ symmetry
	under which the scalars $\phi=\left(\phi_{1},\phi_{2}\right)^{T}$
	and $\phi'=\left(\phi'_{1},\phi'_{2}\right)^{T}$ transform as doublets,
	and $S$ is a scalar singlet. The local operator $\phi\left(x\right)\phi\left(x\right)S\left(x\right)S\left(x\right)$
	is identically zero, but $\phi\left(x\right)\phi\left(y\right)S\left(x\right)S\left(y\right)$
	is not, hence it is possible to collide two $S$'s and create a $\phi_{1}$
	plus a $\phi_{2}$. This is done via two diagrams in which $\phi'_{1}$
	and $\phi'_{2}$ are exchanged in the $t$ and $u$ channels. In the
	limit where the mass $\Lambda$ of $\phi'_{1,2}$ is much larger the
	momentum transferred $p$, the total amplitude varies with $1/\Lambda^{2}\times\left(p/\Lambda\right)^{2}$,
	and not $1/\Lambda^{2}$. }, and that the relevant operator is $\partial\partial LLLLHHHH$ instead
of $LLLLHHHH$.

As pointed out in \cite{Heeck:2013rpa}, the uncertainty in the
invisible decay width of the $Z$ boson as measured indirectly at LEP
\cite{Barate:1999ce,Abreu:2000mh,Acciarri:2000ai,Abbiendi:2000hu}
constrains the mass of the scalar boson in fig.
\eqref{fig:best_diagram} to be heavier than roughly $\Lambda\sim20$
GeV. In this latter process, the momentum $p$ is of the order of the
$Z$ mass, hence $p\sim\Lambda$ and there is no Pauli
blocking. However, the blocking is present in the $0\nu4\beta$ decay
diagram with 4 $W$'s, since $p\sim\textrm{100 MeV}$ is much smaller
than the mass $\Lambda$ of the neutral scalar. As a consequence, the
$0\nu4\beta$ decay lifetime is even larger than what has been assumed
previously \cite{Heeck:2013rpa,Hirsch:2017col}; a simple order of
magnitude calculation yields $\tau_{0\nu4\beta}^{4W}\sim10^{69}$
years, which is very far from the limit given in eq. (\ref{eq:4}).

A possible way of avoiding this particular suppression factor is by
introducing right-handed neutrinos $\nu_{R}$ and $W_{R}$ gauge bosons
such that lepton number is violated by a 4-fermion interaction
$\frac{1}{\Lambda^{2}}\nu_{e,L}\nu_{e,L}\nu_{e,R}\nu_{e,R}$.  This
effective operator can be generated with the scalars $\Delta$ and
$\sigma$ mentioned earlier, keeping in mind that there is now a
$\nu_{e,R}\nu_{e,R}\sigma$ interaction.  Nevertheless, given the
multi-TeV LHC mass limits on the new gauge bosons
\cite{Khachatryan:2016jww,Aaboud:2017efa}, the two diagrams imply a
similar $0\nu4\beta$ decay life-time:
\begin{align}
\tau_{0\nu4\beta}^{2W+2W_{R}}\sim & \left(\frac{g_{R}m_{W_{R}}}{g_{L}m_{W}}\right)^{8}\left(\frac{q}{\Lambda}\right)^{4}\tau_{0\nu4\beta}^{4W}\lesssim\tau_{0\nu4\beta}^{4W}\,.
\end{align}

Finally, we will make some comments about the possible realizations
of the $0\nu4\beta$ decay operator in eq. (\ref{eq:operator_after_EWSB})
with Standard Model fields (see also \cite{Heeck:2014sna}). Due to
Pauli's exclusion principle, fermion fields evaluated at the same
space-time point anti-commute, so for a generic 4-spinor
\begin{align}
\Psi & =\left(\begin{array}{c}
\Psi_{L}^{\uparrow}\\
\Psi_{L}^{\downarrow}\\
\Psi_{R}^{\downarrow}\\
\Psi_{R}^{\uparrow}
\end{array}\right)
\end{align}
we can only have operators of the form $\Psi^{n}\left[\cdots\right]$
with $n\leq4$ (if there are no derivatives).\footnote{If only one
	chirality $X=R,L$ is involved, then
	$\Psi_{X}^{n}\left[\cdots\right]$ with $n\leq2$ are the only
	possibilities. For example, an electron-neutrino mass term
	$\nu_{e,L}\nu_{e,L}$ is allowed, but interactions of the form
	$\nu_{e,L}\nu_{e,L}\nu_{e,L}\left[\cdots\right]$ or
	$\nu_{e,L}\nu_{e,L}\nu_{e,L}\nu_{e,L}$ are not. } Furthermore, if we
expand the Dirac indices of such an operator with $n=4$, the only
non-zero term must be proportional to
$\Psi_{L}^{\uparrow}\Psi_{L}^{\downarrow}\Psi_{R}^{\uparrow}\Psi_{R}^{\downarrow}$.
This means that the Pauli exclusion principle severely restricts local
$0\nu4\beta$ operators to the unique form
\begin{align}
\mathscr{O}_{0\nu4\beta} & =\frac{\kappa}{\sum_{i=1}^{14}\Lambda_{i}}\overline{e}_{L}^{\uparrow}\overline{e}_{L}^{\downarrow}\overline{e}_{R}^{\uparrow}\overline{e}_{R}^{\downarrow}\times\overline{u}\overline{u}\overline{u}\overline{u}dddd\,,
\end{align}
and higher dimensional $0\nu\left(2m\right)\beta$ decay operators,
with $m>2$, are forbidden entirely unless they have derivatives.
Note, that quarks have 3 colors, hence a similar issue arises for
operators with more than 6 quarks of the same charge and chirality. 

In this sense, quadruple beta decay (with or without neutrinos) is
a borderline case between allowed and excluded local processes. An
interesting consequence is that there are only three $0\nu4\beta$
operators of minimal dimension (=18):
\begin{align}
\mathscr{O}_{0\nu4\beta}^{(1)SM} & \sim\overline{L}\overline{L}e^{c}e^{c}\overline{Q}\overline{Q}u^{c}u^{c}\overline{d^{c}}\overline{d^{c}}\overline{d^{c}}\overline{d^{c}}\,,\\
\mathscr{O}_{0\nu4\beta}^{(2)SM} & \sim\overline{L}\overline{L}e^{c}e^{c}\overline{Q}u^{c}u^{c}u^{c}Q\overline{d^{c}}\overline{d^{c}}\overline{d^{c}}\,,\\
\mathscr{O}_{0\nu4\beta}^{(3)SM} & \sim\overline{L}\overline{L}e^{c}e^{c}u^{c}u^{c}u^{c}u^{c}QQ\overline{d^{c}}\overline{d^{c}}\,.
\end{align}
This should be contrasted with the $0\nu2\beta$ decay operators of
dimension 9, of which there are six. On the other hand, the diagram
shown above with 4 $W$'s (fig. \eqref{fig:best_diagram}) and the
equivalent one with $2W$'s + $2W_{R}$'s, have dimensions 24 and 20
respectively:
\begin{align}
\mathscr{O}_{0\nu4\beta}^{\left(4W\right)} & \sim\partial^{2}\overline{Q}\overline{Q}\overline{Q}\overline{Q}QQQQ\overline{L}\overline{L}\overline{L}\overline{L}\overline{H}\overline{H}\overline{H}\overline{H}\,,\label{eq:0Nu4Beta_operator_dim24}\\
\mathscr{O}_{0\nu4\beta}^{\left(2W+2W_{R}\right)} & \sim\overline{Q}\overline{Q}u^{c}u^{c}QQ\overline{d^{c}}\overline{d^{c}}\overline{L}\overline{L}e^{c}e^{c}\overline{H}\overline{H}\,.
\end{align}

\section{\label{sec:OtherLowEnergyProcesses}Other low energy processes with
	$\Delta L\geq4$ involving charged leptons}

We now move on to a brief discussion of other lepton number
violating processes, involving low-energies, with four or more charged leptons. Rough estimates for their rates
will be given in the following. We stress
that for an experimental proof of $L$ violation, final states should not
contain neutrinos.

In table \ref{tab:LowestDimOp} we give the lowest dimension at which a
given $\Delta L\neq0$ operator can appear, together with some examples.
(Note that, since we are interested here in low energy
processes, neither Higgs nor gauge bosons can appear as final
states.) The table starts with the $\Delta L=1$ operators which induce the standard proton decay modes (hence $\Delta B=1$), and these are followed by the $\Delta L=2$ operator associated to neutrinoless double beta decay. With larger $\Delta L$, the dimension of the operators keeps rising and at some point one expects that the rates of associated low-energy processes becomes too small to be observed. We now discuss briefly
this point, by focusing on the most promising processes.

\begin{center}
	\begin{table}[tbph]
		\begin{centering}
			\begin{tabular}{ccc}
				\toprule 
				$\Delta L$ & Minimum operator dimension & Operators with smallest dimension\tabularnewline
				\midrule
				1 & 6 & $e^{1}\overline{d}^{3}$, $e^{1}u^{2}d^{1}$\tabularnewline
				2 & 9 & $e^{2}u^{2}\overline{d}^{2}$\tabularnewline
				3 & 12 & $e^{3}u^{4}\overline{d}$\tabularnewline
				4 & 15 & $e^{4}u^{6}$\tabularnewline
				5 & 21 & $e^{5}u^{6}\overline{d}^{3}$, $e^{5}u^{8}d$\tabularnewline
				6 & 24 & $e^{6}u^{8}\overline{d}^{2}$\tabularnewline
				$\cdots$ & $\cdots$ & $\cdots$\tabularnewline
				\bottomrule
			\end{tabular}
			\par\end{centering}
		
		\protect\caption{\label{tab:LowestDimOp}List of lowest dimensional
			operators, after electroweak symmetry breaking, involving charged
			lepton number violation in $\Delta L$ units. For $\Delta L>4$, more
			than one generation of leptons and/or quarks is necessary (the
			alternative would be to add derivatives). }
	\end{table}
	
	\par\end{center}

It is important to distinguish those cases where baryon number is
violated from those scenarios where where $\Delta B=0$. This is simply
due to the fact that the available energy in $\Delta B \ne 0$
processes is fixed by the nucleon mass, of the order of $\sim$GeV, while
kinetic energy of the charged leptons is much smaller
($\sim$MeV) in the $\Delta B=0$ case.

Let us consider first the latter case, $\Delta B=0$.  This implies
immediately that $\Delta L$ must be an even number.  The relevant
processes are then $0\nu\left(2n\right)\beta$ with $n>2$.  We will
discuss only $\beta^-$ decays, since for quadruple beta decays it has
been shown already in \cite{Heeck:2013rpa} that the positron emission
or electron capture processes are even more hopeless, due to their
smaller Q-values. For $0\nu\left(2n\right)\beta$ with $n>2$ the same
observation applies.

The process $0\nu6\beta$ is induced by an operator with 18 fermions,
hence the decay width $\Gamma$ is suppressed at least by a factor
$Q^{17}q^{29}/\Lambda^{46}$ relative to the nucleon Fermi momentum
$q\sim100$ MeV. Moreover, we note that at most 4 electrons can be at a
single point $x$, therefore the operators for these
$0\nu\left(2n\right)\beta$ decays require at least two derivatives,
and consequently the decay width is suppressed by 4 more powers of
$q/\Lambda$, compared to the simple-minded estimate quoted above.  It
is also straightforward to check that at least 6 electrically charged
bosons are needed to mediate the process hence, the same logic as
discussed for neutrinoless quadruple beta decay applies.

Finally, kinematically $0\nu 6\beta$ and larger is only allowed for
neutron-rich nuclides which are far from the valley of stability,
hence these isotopes will have a very short half-life. The
longest-lived isotope seems to be $_{52}^{134}\textrm{Te}$, which
can decay into $_{58}^{134}\textrm{Ce}+6e^{-}$ with a $Q$ value of 2.3
MeV, but also decays by single beta emission with a half-life of 41.8 minutes
\cite{Audi:2017asy}. For $0\nu8\beta$, considering only isotopes with
an atomic mass below 200, we have
$_{50}^{131}\textrm{Sn}\rightarrow{}_{58}^{131}\textrm{Ce}+8e^{-}$
($Q=2.4$ MeV) and
$_{50}^{132}\textrm{Sn}\rightarrow{}_{58}^{132}\textrm{Ce}+8e^{-}$
($Q=5.9$ MeV) with half-lives
$T_{1/2}\left(_{50}^{131}\textrm{Sn}\right)=56\textrm{ s}$ and
$T_{1/2}\left(_{50}^{131}\textrm{Sn}\right)=39.7\textrm{ s}$. Thus, no
realistic candidate for a $0\nu\left(2n\right)\beta$ experiment with
$n>2$ exists in nature.

Let us now turn to processes where baryons are destroyed and hence,
the available energy is much larger.  Here, we consider the cases
$\left(\Delta L,\Delta B\right)=\left(4,\pm2\right)$ and $\left(\Delta
L,\Delta B\right)=\left(5,\pm1\right)$; violation of lepton or
baryon number in greater quantities will be associated with even
bigger minimum life-times.

The lowest dimensional operator with 4 charged leptons (after
electroweak symmetry breaking) is $eeeeuuuuuu/\Lambda^{11}$ (see table
\ref{tab:LowestDimOp}).  It leads, for example, to diproton decay:
$\left(A,Z\right)\rightarrow\left(A-2,Z-2\right)+2\pi^{-}+4e^{+}$.
In the most optimistic scenario, this operator can be built in such a
way that only 7 powers of $\Lambda$ correspond to the mass of
mediators with colour ($\Lambda_{C}$), while the remaining 4 powers of
$\Lambda$ are related to the mass of fields with electroweak
interactions only ($\Lambda_{EW}$ ). Hence,
\begin{align}
\tau_{min}^{-1} & \sim10^{-13}\frac{\left(2m_{p}\right)^{23}}{\Lambda_{EW}^{8}\Lambda_{C}^{14}}\sim10^{-40}\textrm{y}^{-1}\,,
\end{align}
using the values $\Lambda_{EW}\approx200$ GeV and
$\Lambda_{C}\approx2$ TeV. The prefactor $10^{-13}$ takes care of the
fact that this is a 6-body decay. Super-Kamiokande has searched for
other diproton decay modes, imposing limits of the order of $10^{32}$
years on the associated life-times \cite{Gustafson:2015qyo}.  It seems
therefore very hard to probe $\Delta L=4$ processes at low energies,
even for those cases where baryon number is violated. Note, however,
that due to the much larger energy release, the gap between the
experimental sensitivity and the most optimistic expectation is only 8
orders of magnitude, compared to the (minimum of) 20 orders found for
quadruple beta decay.

For $\Delta L=5$ (or larger) the decay rates are necessarily even more
suppressed.  For $\Delta L=5$, the lowest dimensional operators have
$d=21$, as shown in table \ref{tab:LowestDimOp}, therefore the decay
widths will be suppressed by 34 powers of $\sim m_{p}/\Lambda$ when
compared to $m_{p}$.  In summary, we conclude that observation of
charged lepton number violation in four or more units in low energy
experiments will be impossible in the foreseeable future.

\section{\label{sec:Colliders}Probing lepton number violation at colliders}

We now turn to a discussion of probing models with multiple lepton
number violation at colliders. We do not aim at a full, systematic
analysis of all possibilities. Instead, we will discuss two simple
models with $\Delta L=4$ and then present one example for a model with
$\Delta L=5$. Models which lead to different $\Delta L=4$ operators at
low energy or models with larger $\Delta L$ violation can be easily
constructed following the same principles that we use in our examples.

Our first model is inspired by the discussion on neutrinoless
quadruple beta decay in section \ref{sec:0Nu4Beta}. In this model,
called model-I  below, we add only two new fields to the standard
model. Both are scalars: (i) $\Delta = S_{1,3,1,-2}$ and (ii)
$T=S_{1,3,0,-2}$. Here, the subscripts stand for the transformation
properties or charge under the SM gauge group and lepton number,
$SU(3)_C \times SU(2)_L \times U(1)_Y,L$ (also we use an $S$ for
scalars and, later on, an $F$ for 2-component Weyl spinors). The only
change with respect to the model discussed in section
\ref{sec:0Nu4Beta} is that we have replaced the singlet
$\sigma=S_{1,1,0,-2}$ with the $Y=0$ $SU(2)_L$ triplet field $T$.

As before, we enforce a $Z_4(L)$ symmetry which ensures that leptons
can only be created or destroyed in groups of 4.
The Lagrangian of the model is:\footnote{We use the notation
	$\Delta\equiv\left(\begin{array}{cc} \Delta^{++} &
	\frac{\Delta^{+}}{\sqrt{2}}\\ \frac{\Delta^{+}}{\sqrt{2}} &
	\Delta^{0}
	\end{array}\right)$, $T\equiv\left(\begin{array}{cc}
	T^{+} & \frac{T^{0}}{\sqrt{2}}\\
	\frac{T^{0}}{\sqrt{2}} & T^{-}
	\end{array}\right)$ and $\epsilon\equiv\left(\begin{array}{cc}
	0 & 1\\
	-1 & 0
	\end{array}\right)$.} 
\begin{eqnarray}\label{eq:lag1}
{\cal L} &=& {\cal L}^{\rm SM}  + \left[Y_{\Delta} \left(L^T\epsilon \Delta \epsilon L\right)
- \lambda_{HH\Delta T} \left(H^T\Delta^*T\epsilon H\right) + \textrm{h.c.}\right] 
+ m_{\Delta}^2 \textrm{Tr}(\Delta^{\dagger}\Delta) + m_T^2 \textrm{Tr}\left(T\epsilon T \epsilon\right) \\ \nonumber 
&+&  \lambda_{T} \left[\textrm{Tr}\left(T\epsilon T \epsilon\right)\right]^2
-  \lambda_{HT} (H^\dagger H) \textrm{Tr}\left(T\epsilon T \epsilon\right) 
-  \lambda_{\Delta1} \left[\textrm{Tr}(\Delta^{\dagger}\Delta)\right]^2
-  \lambda_{\Delta2} \textrm{Tr}(\Delta^{\dagger}\Delta\Delta^{\dagger}\Delta)
\\ \nonumber
&-&  \lambda_{\Delta H1} (H^\dagger H) \textrm{Tr}(\Delta^{\dagger}\Delta)
-  \lambda_{\Delta H2}  \left(H^\dagger \Delta \Delta^{\dagger} H \right)
-  \lambda_{\Delta T1} \textrm{Tr}(\Delta^{\dagger}\Delta) \textrm{Tr}\left(T\epsilon T \epsilon\right)
-  \lambda_{\Delta T2} \textrm{Tr}(\Delta^{\dagger} T\Delta T)
\end{eqnarray}

Note that this Lagrangian is also $U(1)_B$ invariant.  In other words,
baryon number is preserved, hence processes such as dinucleon decay
are completely absent in this model.  Note that $\Delta$ is the same
field that appears in the type-II seesaw mechanism. However, our
symmetry forbids the term $H\Delta H$, which in seesaw type-II is the
source of $\Delta L=2$. This implies that for $m_{\Delta}^2 \ge 0$, in
our model there is no induced vev for $\Delta^0$ and thus no Majorana
neutrino mass term.

We have not written generation indices in eq. (\ref{eq:lag1}).  In
general $Y_{\Delta}$ is a complex symmetric (3,3) matrix. All terms in
the Lagrangian, with the exception of those proportional to $T^2$ or
$T^4$, conserve lepton number. For the phenomenology discussed below it
is important that $m_T^2$ violates $\Delta L$ in 4 units.

The term proportional to $\lambda_{HH\Delta T}$ leads to mixing between the
neutral and singly charged components in $T$ and $\Delta$ after
electro-weak symmetry breaking (EWSB), as well as to a mass splitting
between the CP-even and CP-odd components in $\Delta^0$. Thus, after
EWSB the model has two new neutral CP-even scalars $S^0_{1,2}$ and two
singly charged scalars $S^{\pm}_{1,2}$ plus one neutral CP-odd scalar,
$A^0$, and one doubly charged scalar, $\Delta^{\pm\pm}$.  In our
numerical calculations we always diagonalize all mass matrices and
consider mass eigenstates correctly.  However, in the following
discussion we simply use $T^0$ and $\Delta^0$ for $S^0_2$ and $S^0_1$,
respectively.  (And similarly for the singly charged states $T^{\pm}$
and $\Delta^{\pm}$.)  This is done only for the clarity of the
discussion; it does not affect any of our conclusions. Note that for
typical choices of masses $m_T^2$ and $m_{\Delta}^2$ above
$(500\hskip1mm {\rm GeV})^2$ mixing between the different states will
be small unless $m_T^2 \simeq m_{\Delta}^2$.

For the numerical calculations shown below, we have implemented the
model into SARAH \cite{Staub:2012pb,Staub:2013tta}. The implementation
is then used to generate SPheno code \cite{Porod:2003um,Porod:2011nf}
for the numerical generation of spectra. The UFO model files generated
by SARAH are used for cross section and decay calculations with
MadGraph \cite{Alwall:2014hca}.

\begin{figure}[H]
	\begin{centering}
		\includegraphics[scale=0.5]{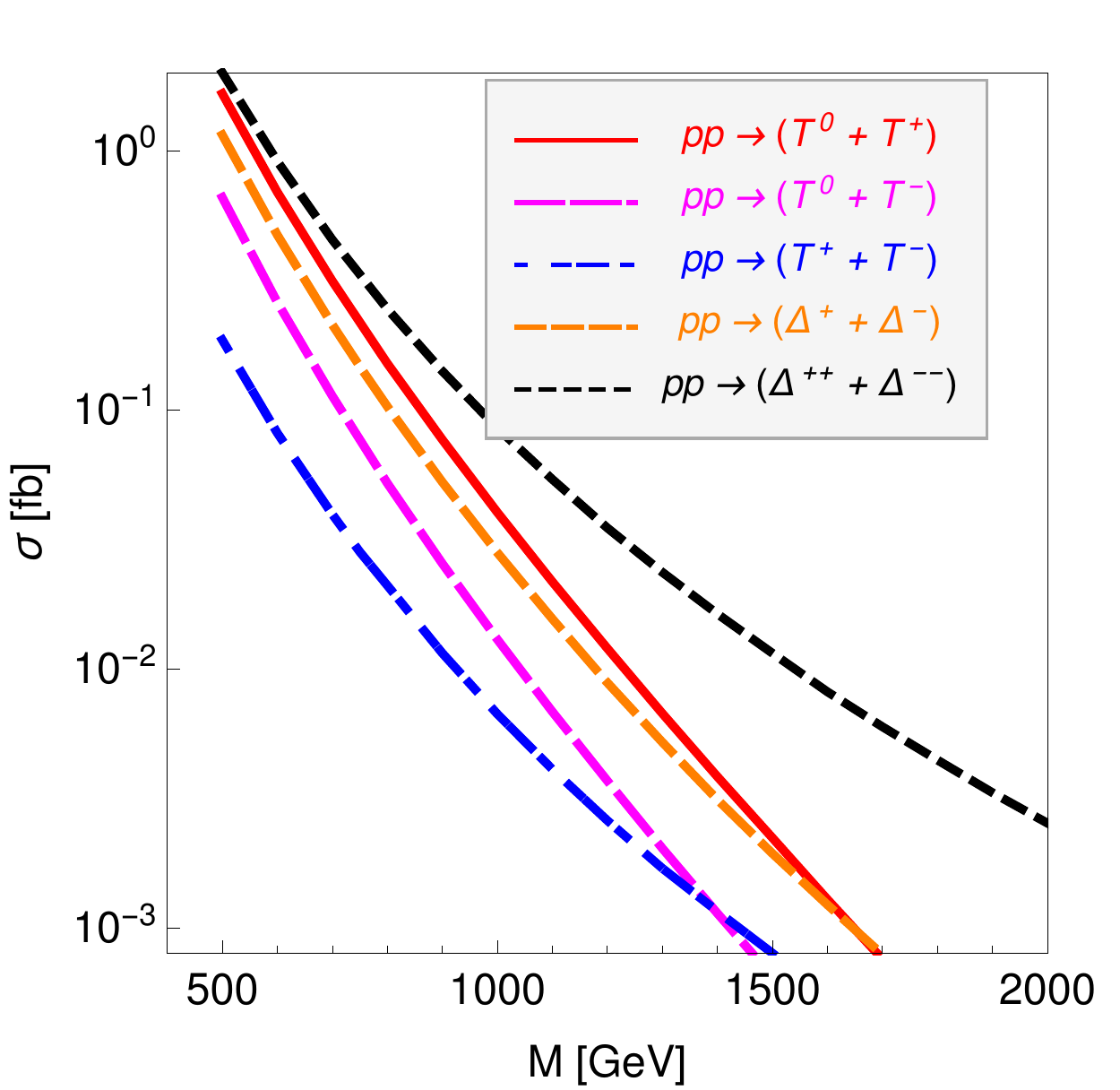}\hskip5mm
		\includegraphics[scale=0.5]{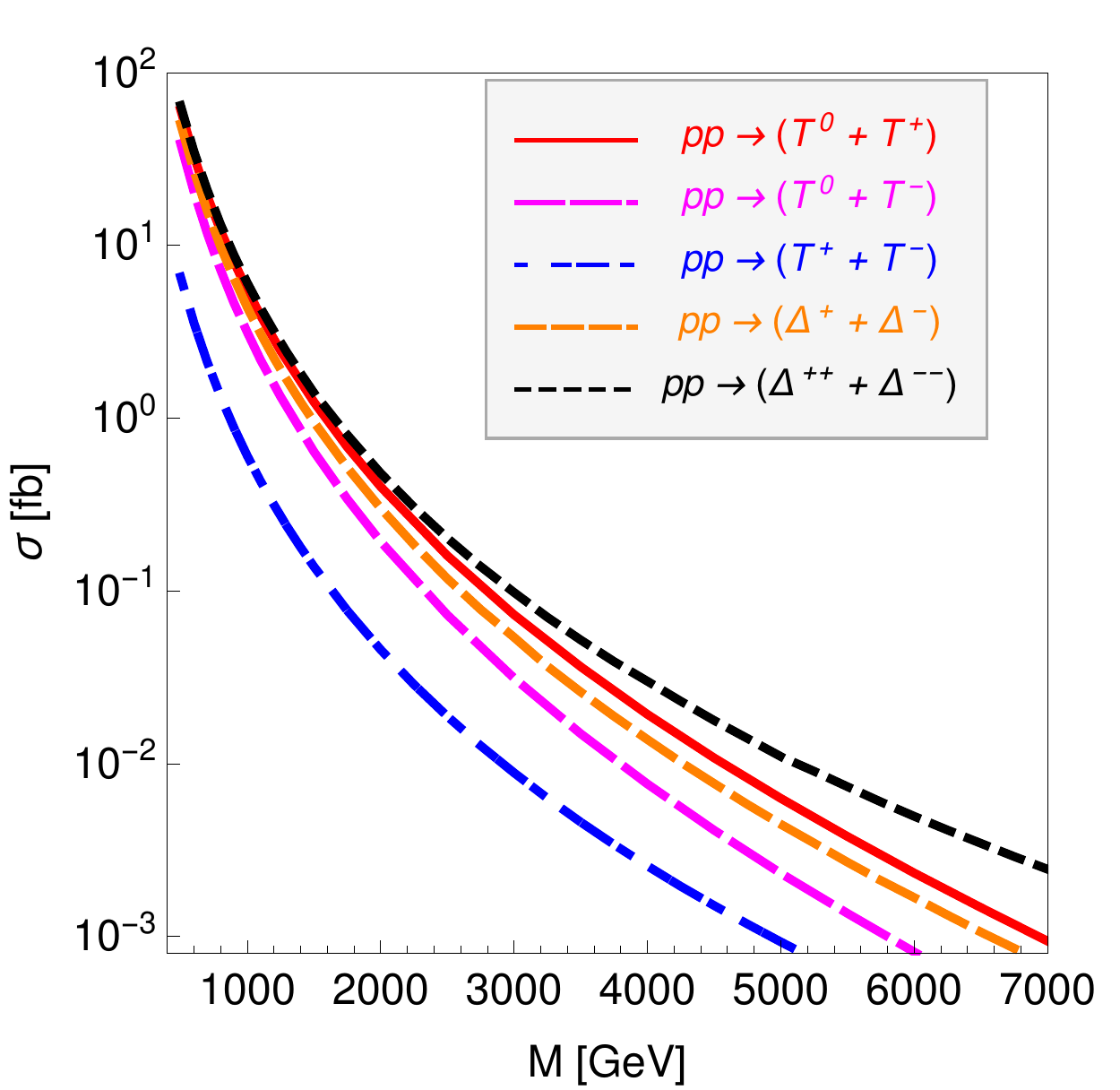}
		\par\end{centering}
	
	\protect\caption{\label{fig:XsecMI} Cross sections for various
		production modes for particles in our $\Delta L=4$ model I. The
		plot to the left is calculated for the LHC, i.e. pp collisions at
		$\sqrt{s}=13$ TeV. The plot on the the right assumes a hypothetical
		future pp-collider with $\sqrt{s}=100$ TeV. Note the different
		scales in the two plots.}
\end{figure}

\begin{figure}[H]
	\begin{centering}
		\includegraphics[scale=1.2]{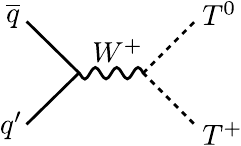}
		\vskip5mm
		\includegraphics[scale=1.2]{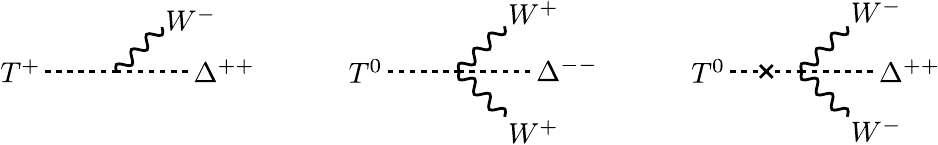}
		\par\end{centering}
	
	\protect\caption{\label{fig:DiagsMI} Feynman diagrams for the
		production and decays of $T^0$ and $T^{\pm}$.  This list of diagrams
		is not complete, showing instead only those diagrams relevant for
		the observation of a $\Delta L=4$ process, see text.}
\end{figure}

Both, $\Delta$ and $T$ can be produced with sizeable rates at
colliders.  Fig. (\ref{fig:XsecMI}) shows cross sections for the most
important production modes in pp-colliders for two values of
$\sqrt{s}$: To the left $\sqrt{s}=13$ TeV, to the right $\sqrt{s}=100$
TeV. The numerically largest cross section is found for
$\Delta^{\pm\pm}$ pair production.  However, as we will discuss below,
for the observation of $\Delta L=4$ processes the interesting
production modes are $pp \to (T^0 + T^+)$ and $pp \to (T^0+T^-)$. Both
processes proceed through an off-shell $W$-diagram, see
fig. (\ref{fig:DiagsMI}). The cross section for $pp \to (T^0+T^+)$ is
larger than for $pp \to (T^0+T^-)$, reflecting the fact that the
initial state is positively charged.

We first discuss the decays of $\Delta$. The different components of
$\Delta$ decay according to $\Delta^{\pm\pm} \to
l^{\pm}_{\alpha}l^{\pm}_{\beta}$, $\Delta^{\pm} \to
\nu_{\alpha}l^{\pm}_{\beta}$ and $\Delta^{0} \to
\nu_{\alpha}\nu_{\beta}$ with 100\% branching ratio, when summed over $\alpha$ and
$\beta$. Since cross sections are largest and background lowest for
$\Delta^{\pm\pm}$, the most stringent constraints on $m_{\Delta}$
come from searches for $\Delta^{\pm\pm}$.  Both ATLAS
\cite{ATLAS:2017iqw} and CMS \cite{CMS:2017pet} have searched for
doubly charged scalars decaying to charged leptons.  Limits depend
quite strongly on the flavour of the charged leptons.  CMS
\cite{CMS:2017pet} gives limits as low as $m_{\Delta^{\pm\pm}} \simeq
535$ GeV for a $\Delta^{++}$ decaying with 100 \% to pairs of taus,
while limits are in the range of ($800-820$) GeV, if the $\Delta^{++}$
decays only to electrons or muons.  ATLAS \cite{ATLAS:2017iqw}, on the
other hand, has established lower limits on $m_{\Delta^{\pm\pm}}$ of
roughly ($600-800$) GeV, for branching ratios to either electrons or
muons in the range of ($0.2-1$). We will therefore use two choices of
$m_{\Delta}$ in our numerical examples below, namely, $m_{\Delta}=0.6$
TeV and $m_{\Delta}=1$ TeV. The former is allowed only for a $\Delta$
with coupling mostly to $\tau$'s, while the latter is currently
unconstrained. Note, however, that with the predicted ${\cal L}=3/$ab
for the high luminosity LHC (HL-LHC) $m_{\Delta}$ in excess of
$m_{\Delta}=1$ TeV will be probed, while for a $\sqrt{s}=100$ TeV collider
we estimate from fig. (\ref{fig:XsecMI}) that up to $m_{\Delta} \sim 5$ TeV
could be tested with ${\cal L}=3/$ab, in agreement with the numbers
quoted in \cite{Arkani-Hamed:2015vfh}.

\begin{figure}[H]
	\begin{centering}
		\includegraphics[scale=0.6]{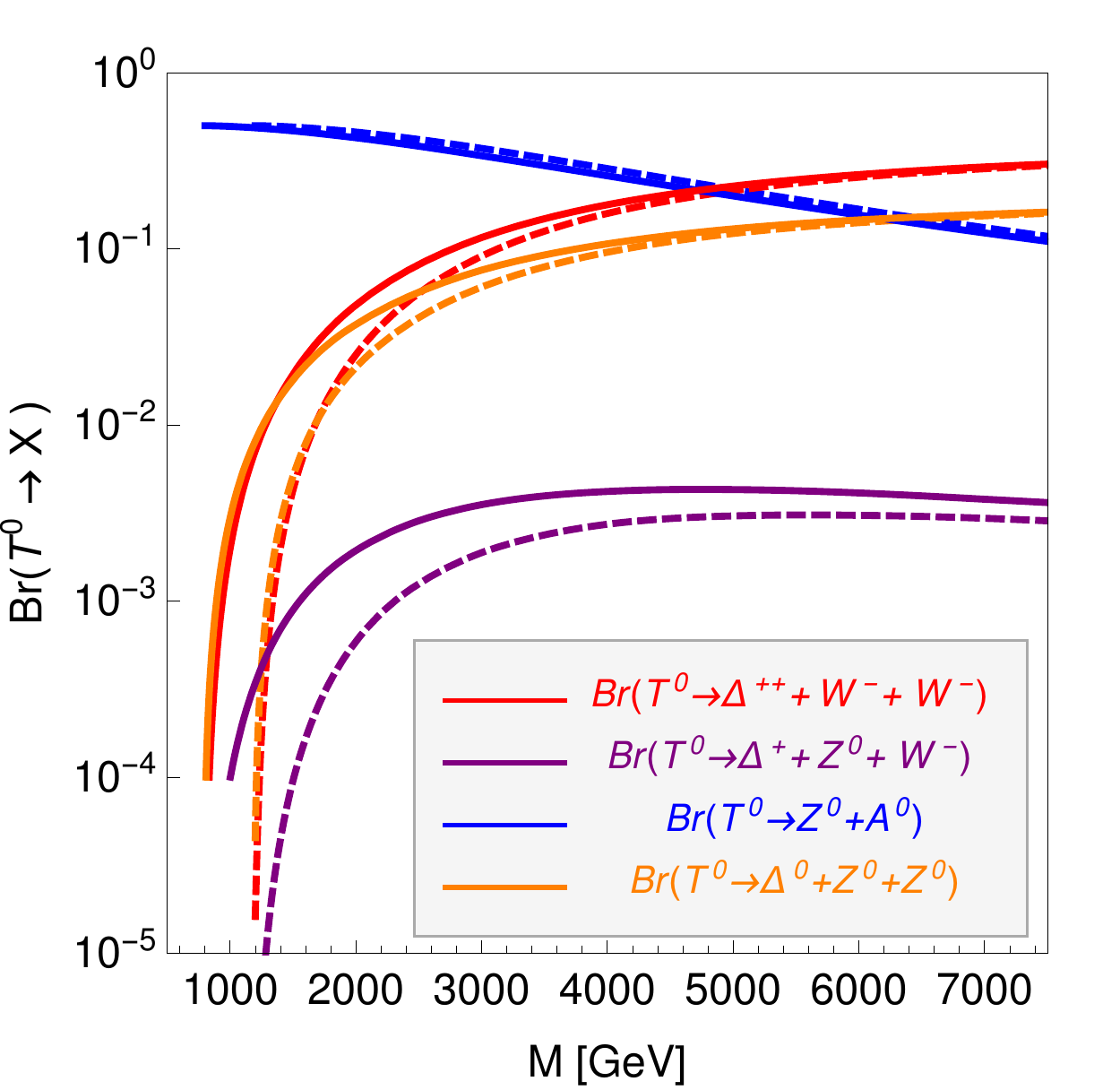}
		\par\end{centering}
	
	\protect\caption{\label{fig:BrT0} Branching ratios for the decay of
		$T^0$ as function of the mass of $T^0$. Here, the full lines are for
		the choice $m_{\Delta}=0.6$ TeV, the dashed lines for $m_{\Delta}=1$
		TeV. In addition to $T^0 \to A^0 +Z^0$ there is a second two-body
		decay mode for $T^0$, $T^0 \to h +\Delta^0$. This one is not shown
		explicitly, since Br($T^0 \to h +\Delta^0$) $\simeq$ Br($T^0 \to A^0
		+Z^0$) in all cases. Note that Br($T^0 \to \Delta^{++}+2W^-$)
		=Br($T^0 \to \Delta^{--}+2W^+$). The coupling $\lambda_{HH\Delta T}$
		was chosen $\lambda_{HH\Delta T}=0.1$ in this example, while the entries
		in $Y_{\Delta}$ where arbitrarily put to be smaller than ${\cal
			O}(0.1)$. With these choices, decays to purely leptonic final
		states are negligible and therefore not shown.}
\end{figure}

For the observation of a $\Delta L=4$ processes, we need to produce
the hyperchargeless triplet $T$. We therefore now turn to a discussion
of the decays of $T^0$ and $T^{\pm}$.  First of all, note that all
decay rates for these particles will be proportional to the coupling
$\lambda_{HH\Delta T}$. This is due to the fact that this term is the only
one linear in $T$ allowed by $Z(4)_L$.  Mixing between $T^0$ and
$T^{\pm}$ with $\Delta^0$ and $\Delta^{\pm}$ induce two-body decays
for these states into leptonic final states.  However, these will
always involve neutrinos and thus are not useful to establish
experimentally lepton number violation.  More important are then
decays of $T^0$ and $T^{\pm}$ to $\Delta^{\pm\pm}$ and gauge
bosons. Fig. (\ref{fig:DiagsMI}) shows the most important Feynman
diagrams. Apart from $T^{+}\to W^-+ \Delta^{++}$, $T^+$ can decay to
$W^+ + A^0$, $W^+ + \Delta^0$,\footnote{Recall that we use $\Delta^0$
	in this discussion synonymous for the CP-even scalar $S^0_1$.}  as
well as $\Delta^+ + h$ and $\Delta^+ + Z^0$. The branching ratio for
$T^{+}\to W^-+ \Delta^{++}$ is always close to 25\%, if the mixing
between $T^{+}$ and $\Delta^+$ is small. Similarly, $T^0$ can decay to
a number of final states involving gauge (and Higgs) bosons.
Fig. (\ref{fig:BrT0}) shows the most important branching ratios for
$T^0$ decays as a function of $m_T$. As the plot shows, for small
values of $m_T$ the two-body decays $T^0 \to A^0 + Z^0$ (and $T^0 \to
\Delta^0 + h$) have the largest branching ratios. However, for large
values of $m_T$, the decay mode $T^0 \to \Delta^{\pm\pm}+2W^{\mp}$
becomes dominant. $T^0$ decays to both $\Delta^{++}$ and $\Delta^{--}$
with equal rates. That both decays have the same rate can be
understood as a mass insertion in the decay of $T^0$, see
fig. (\ref{fig:DiagsMI}).  Recall that $m_T^2$ violates $L$ by 4
units.

\begin{figure}[H]
	\begin{centering}
		\includegraphics[scale=0.6]{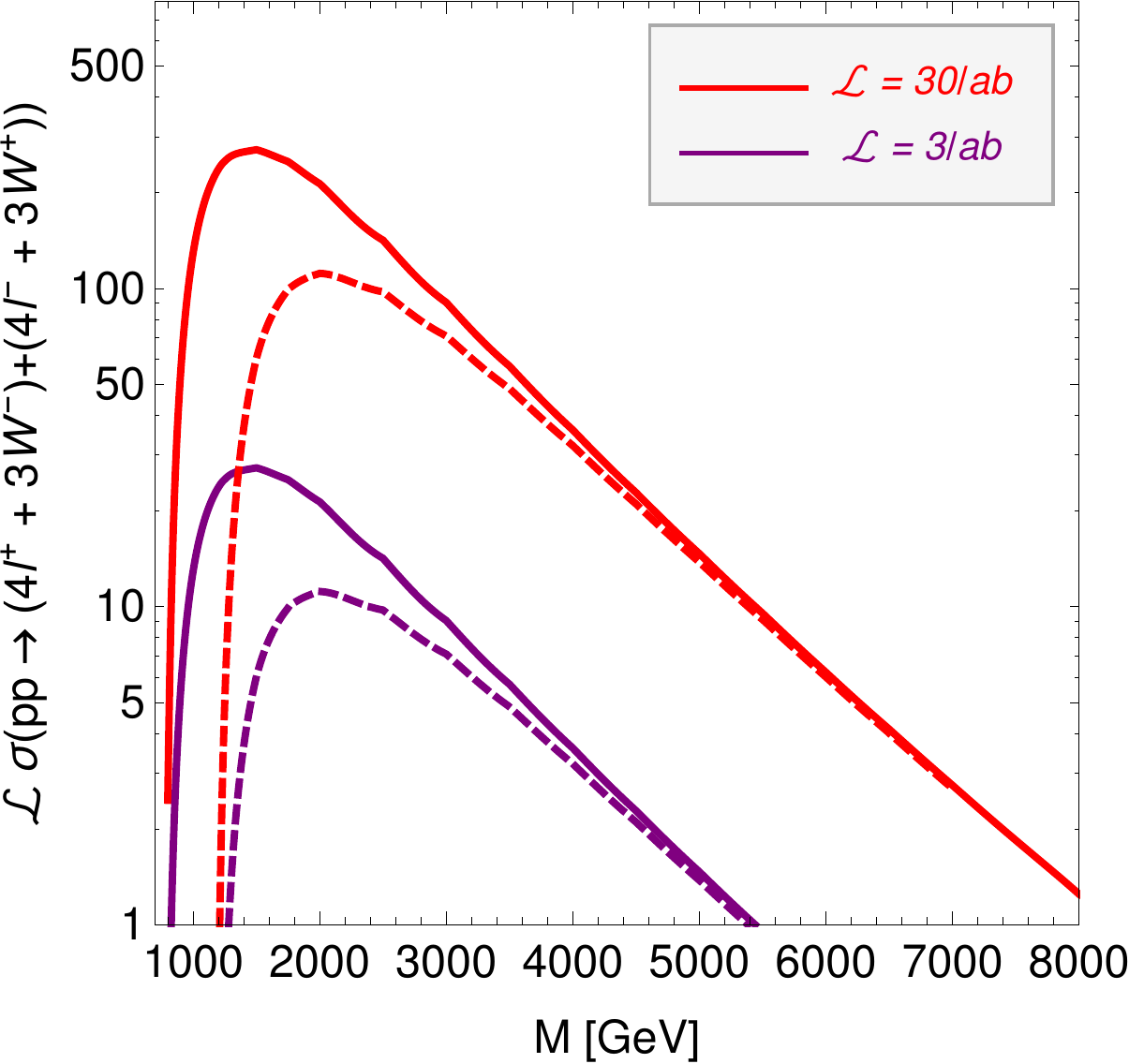}
		\par\end{centering}
	
	\protect\caption{\label{fig:Del4} Maximal number of $\Delta L=4$
		events attainable in model-I at a future pp-collider with
		$\sqrt{s}=100$ TeV. Here, the event number sums over both,
		negatively and positively charged leptons.}
\end{figure}

At a pp collider one can therefore have lepton number violating
(conserving) final states from the chain $pp \to T^+ + T^0 \to 2
\Delta^{++}+3 W^- \to 4 l^+ + 6 j$ ($pp \to T^+ + T^0 \to
\Delta^{++}+\Delta^{--}+2 W^++W^- \to 2 l^+ 2 l^- + 6j$).  With the
cross sections, see fig. (\ref{fig:XsecMI}), and the decay branching
ratios we can then calculate the number of $\Delta L=4$ events as a
function of $m_T$ (and $m_{\Delta}$).  We have performed this exercise
for the LHC and find that, taking into account the lower limit on
$m_{\Delta}$, there will be less than 1 event even after $3/ab$ of
data has been taken. However, the prospects look much brighter at an
hypothetical $\sqrt{s}=100$ TeV collider, see fig. (\ref{fig:Del4}).
In this figure we show the maximal number of $\Delta L=4$ events
attainable in model-I for two choices of $m_{\Delta}$ and two different
values for the expected luminosity. For ${\cal L}=30/$ab more than 10
$\Delta L=4$ events could be found up to roughly $m_T \simeq 6$ TeV.

Thus, for our first model we conclude that while the LHC can extend
the search for the different particles in model-I to above 1 TeV,
observation of $\Delta L=4$ seems not to be possible at the LHC.
At a $\sqrt{s}=100$ TeV collider $\Delta L=4$ could be discovered
in this model up to a scale of roughly 6 TeV.

The negative conclusions for the LHC can be simply understood from the
fact that model-I contains no new coloured states.
As shown in table \eqref{tab:LowestDimOp}, the smallest dimensional
operator generating four charged leptons is 15-dimensional: $e^4u^6$.
We therefore choose to implement it in our second example, model-II.
Any model leading to this operator will necessarily involve
beyond-the-SM coloured fields.

Model-II introduces three new states, two fermions $O=F_{8,1,0,-2}$
and $D^c=F_{{\bar 3},1,1/3,1}$ (together with its vector-partner
${\bar D^c}=F_{3,1,-1/3,3}$) and one scalar, $S_d = S_{3,1,-1/3,1}$.
By enforcing once again $Z_4(L)$ invariance, we obtain this time a
Lagrangian with an enlarged accidental symmetry group $G_{SM}\times
U(1)_{2B-L}$. It might not be immediately obvious that this latter
group contains $Z_4(L)$, but this is nevertheless true. Indeed, the
Lorentz and Standard Model group $G_{SM}$ force all operators with SM
fields to be $Z_2(B-L)$ invariant, hence we may write $L=2n+B$ for
some integer $n$. Together with $L=2B$, it is then quite easy to see
that $L$ and $B$ are forced to be multiples of 4 and 2,
respectively. Crucially, unlike in model-I, due to the $U(1)_{2B-L}$
symmetry it is not possible to break lepton number without breaking
baryon number as well. As such, one can have small/unobservable
dinucleon decays rates, but neutrinoless quadruple beta decay is
strictly forbidden.

\begin{figure}[H]
	\begin{centering}
		\includegraphics[scale=1.2]{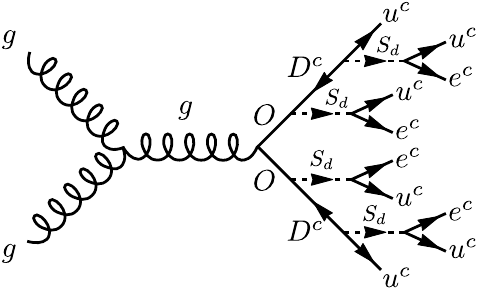}
		\par\end{centering}
	
	\protect\caption{\label{fig:M2} Example diagram for production and
		decays of the color octet in model-II.}
\end{figure}

The Lagrangian contains the following terms:
\begin{eqnarray}\label{eq:lagII}
{\cal L} &\propto& Y_1 u^c e^c S_d^{\dagger} + Y_ 2 u^c D^c S_d
+ Y_3 Q L S_d + Y_ 4 {\overline D^c}O S_d + Y_ 5 D^c O S_d^{\dagger}
+ h.c. \\ \nonumber
&+& m_O O O + m_D {\overline D^c} D^c.
\end{eqnarray}
We have calculated the pair production cross sections for the new
particles in our model-II using again MadGraph \cite{Alwall:2014hca}.  The
results are shown in fig. (\ref{fig:XsecMI}). Again, the plot to the
left is for the LHC, the one on the right is calculated for
$\sqrt{s}=100$ TeV. In all cases one expects that gluon-gluon fusion
gives the largest contribution to the cross section, see
fig. (\ref{fig:M2}). The largest cross section is found for the
fermionic octet. More than 10 events from $O$-pair production are
expected in ${\cal L}=3$/ab for octet masses up to $m_O \simeq 3$
TeV. A $\sqrt{s}=100$ TeV collider would be able to collect more than
10 events for octet masses up to $m_O \simeq 15.5$ ($18.5$) TeV for
${\cal L} =3/$ab ($30/$ab).

\begin{figure}[H]
	\begin{centering}
		\includegraphics[scale=0.5]{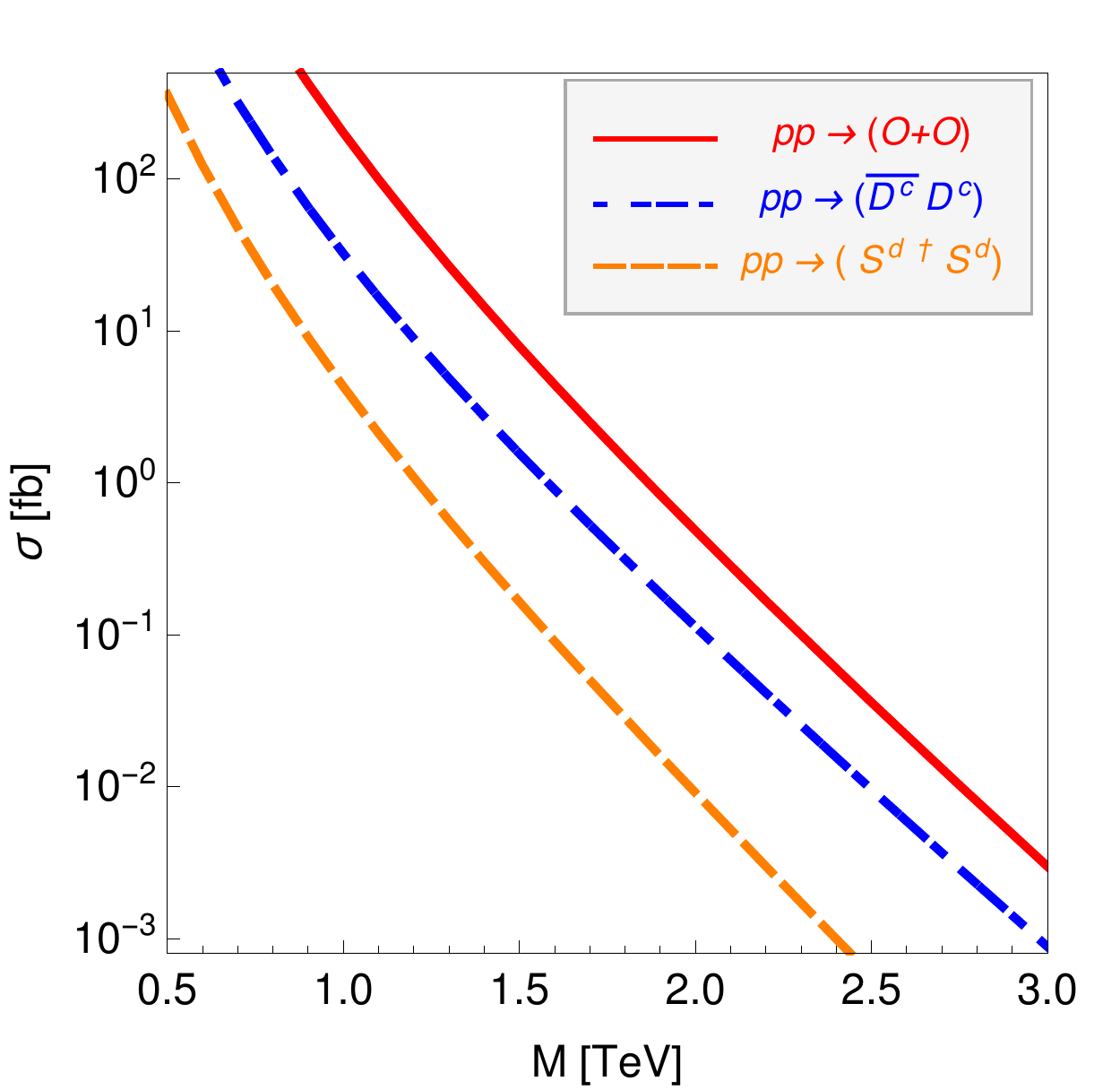}\hskip5mm
		\includegraphics[scale=0.5]{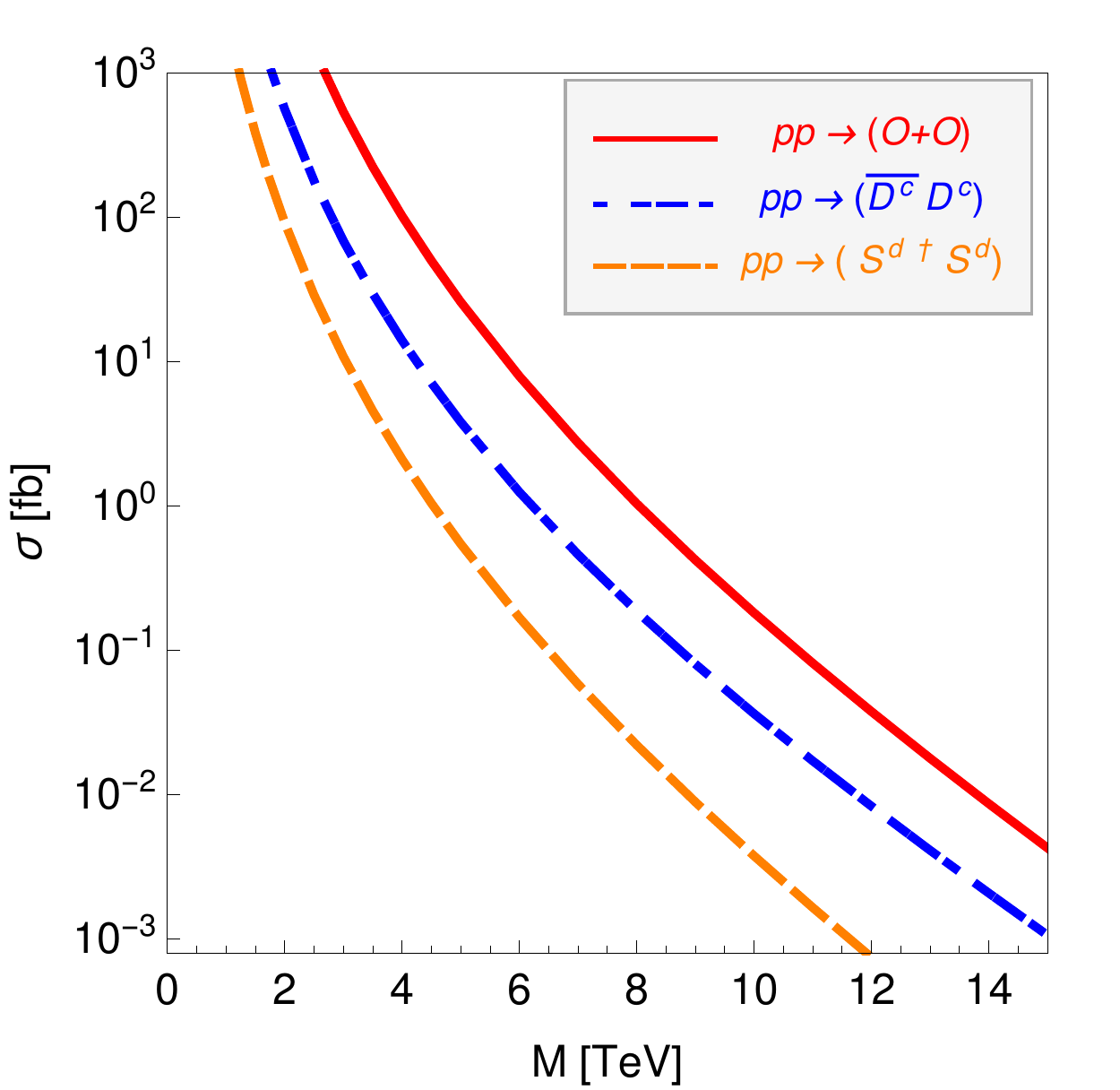}
		\par\end{centering}
	
	\protect\caption{\label{fig:XsecMII} Cross sections for various
		production modes for particles in our $\Delta L=4$ model II. The
		plot to the left is calculated for $\sqrt{s}=13$ TeV, the one on the
		right for $\sqrt{s}=100$ TeV. }
\end{figure}

In model-II, since $O$ is an electrically neutral state, it will decay
with equal branching ratios to $ D^c_{1/3}+ S_{d,-1/3}$ and
$D^c_{-1/3} +S_{d,1/3}$. This implies that $\Delta L=4$ final states
$4 l^+ + 6 j$ will have the same rate as the $\Delta L=0$ final states
$2 l^++2 l^- + 6 j$. Thus, the production cross section (event number)
of $O$-pair production gives directly the limit on the scale up to
which $\Delta L=4$ can be tested in model-II.

\begin{figure}[H]
	\begin{centering}
		\includegraphics[scale=1.2]{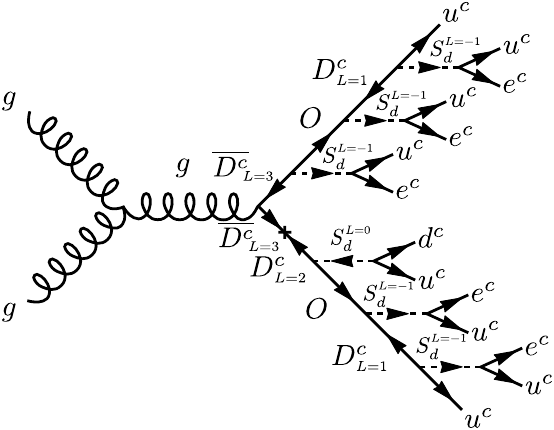}
		\par\end{centering}
	
	\protect\caption{\label{fig:M3}
		Example diagram for a $\Delta L=5$ process in model-III.}
\end{figure}

It is straight-forward to use the ideas discussed above to construct
also models, which lead necessarily to larger $\Delta L$.  We will
discuss only one example with $\Delta L=5$. This model-III introduces
five new states. We will need two copies of $S_d$, to which we assign
different lepton numbers, $S_d^{L=1} = S_{3,1,-1/3,1}$ and $S_d^{L=0}
=S_{3,1,-1/3,0}$. The vector-like down quarks also now come in two
copies. We have $D^c_{L=2}=F_{{\bar 3},1,1/3,2}$, its vector partner
is ${\bar D^c}_{L=3}=F_{3,1,-1/3,3}$, and also $D^c_{L=1}=F_{{\bar
		3},1,1/3,1}$ with its vector partner ${\bar
	D^c}_{L=4}=F_{3,1,-1/3,4}$. Finally, the model also has the
fermionic vector octet, $O=F_{8,1,0,2}$ and ${\bar O}=F_{8,1,0,3}$.
With this lepton number charges, we then enforce a $Z_5(L)$ symmetry.
Just as with model-II, there is a bigger, accidental symmetry group in
this model, $U(1)_{5B-3L}$. In other words, for each group of 5
leptons created, 3 new baryons should appear as well, and for this
reason the proton is completely stable in this model-III.

$\Delta L=5$ processes at the LHC can then occur through diagrams such
as the example shown in fig. (\ref{fig:M3}), where ${\bar D^c}_{L=3}$
is pair produced via gluon fusion.  Note that the decay chains of both
${\bar D^c}_{L=3}$ and $\left({\bar D^c}_{L=3}\right)^*$ end with the
same number of SM fermions: Seven. One can assign the source of
lepton number violation in this diagram to the mass term
${\bar D^c}_{L=3} D^c_{L=2}$.
If all other particles in the diagram are lighter than $D^c_{L=2}$ and
all couplings the same order, $\Delta L=5$ and $\Delta L=0$ final
states from these decay chains will have similar rates.\footnote{If
	all masses and all couplings are numerically the same, the branching
	ratio for $\Delta L=5$ and $\Delta L=0$ final states becomes equal.}
Nevertheless, note that even if the ${\bar D^c}_{L=3} D^c_{L=2}$ mass
term was switched off, lepton number would still be broken; in fact,
$B$ and $L$ conservation would be restored only if the vector masses
of $D^c_{L=1}$ and $O$ were absent as well.

We can estimate the mass reach of the LHC to test this kind of diagram
from the cross sections shown in fig. (\ref{fig:XsecMII}). We estimate
that more than 10 events (before cuts) would remain for masses of
$D^c_{L=2}$ below 2.7 TeV in ${\cal L}=3/$ab. At a $\sqrt{s}=100$ TeV
collider more than 10 events would occur for $m_{D^c}$ below 13.3 (15.5)
TeV in ${\cal L}=3/$ab ($30/$ab). Thus, there is amble parameter space
that could be tested in future colliders even for models with
$\Delta L > 4.$. Many different models of this kind can be readily
constructed.

\section{\label{sec:Summary}Summary}

Given the current experimental situation, the total number of leptons
$L$ might be a conserved quantity. Standard probes for $\Delta L\ne 0$
are proton decay ($\Delta L=1,3$) and neutrinoless double beta decay
($\Delta L=2$) experiments.  However, neither have found any signal so
far.  It is therefore possible that $L$ is violated only in larger
multiplicities, i.e. three, four or more units.

In this context, we have discussed that the decay of a nucleus into
four electrons and no neutrinos will likely never be observed.  This
and other $\Delta L=4$ low energy processes where the leptons are
electrically charged must necessarily be mediated by several heavy
bosons, hence their amplitudes are severely suppressed, given the
current accelerator constraints. We have calculated a very
conservative lower limit on the half-life for neutrinoless quadruple beta decay,
which is 20 orders of magnitude larger than current experimental
sensitivities.  The same conclusion is valid if lepton number is broken
in five or more units. The least pessimistic scenario that we have
found is the one where baryon number is also violated. In this case
two protons in a nucleus could decay into 4 positrons plus pions, for
example. The energy scale in this process is set by the proton mass,
but even so, the high dimensionality of the operators involved implies
that proton decay experiments would need to increase their exposure by
at least 8 orders of magnitude before meaningful constraints could be
derived experimentally.

Colliders, on the other hand, can explore the possibility that lepton
number is violated in four or more units. The reason behind this
observation is rather simple: Even though these processes involve many
exotic particles, if the energy of the collider exceeds the mass of
those exotics, the suppression associated with the high dimensionality
of these $\Delta L\geq4$ operators disappears.  In this work, we
presented two models for $\Delta L=4$ and one for $\Delta L=5$, which
make use of this idea. We have calculated cross sections for the LHC
and a possible future $\sqrt{s}=100$ TeV collider, estimating the
rates for $\Delta L=4$ (and $\Delta L=5$) processes.  Naturally, for
hadron colliders the expectations are highest for models which
contain coloured particles. In this case the LHC (the $\sqrt{s}=100$
TeV collider) could probe $\Delta L=4$ up to scales of roughly 3 TeV
(18 TeV). We expect that the high multiplicity of these events
associated with $\Delta L \geq 4$ should make them virtually
background free, giving a rather spectacular signal.

\section*{Acknowledgments}

This work was funded by the Spanish state through the projects FPA2017--85216--P
and SEV--2014--0398 (from the \textit{Ministerio de Econom\'ia, Industria
	y Competitividad}), as well as PROMETEOII/2014/084 (from the \textit{Generalitat
	Valenciana}). R.F. was also financially supported through the grant
\textit{Juan de la Cierva-formaci\'on} FJCI-2014-21651.

\end{document}